\documentclass{article}

\usepackage{amsmath}  
\usepackage{amsfonts} 
\usepackage{graphicx} 
\usepackage{textcomp}
\usepackage{indentfirst}
\usepackage[authoryear]{natbib}
\bibpunct{(}{)}{;}{a}{}{,}
\begin{document}


\title{Origin of magnetars in binary systems}

\author{Sergei B. Popov}



\maketitle 

\begin{center}
Sternberg Astronomical Institute, Lomonosov Moscow State University,\\
Russia, Moscow, 119991, Universitetsky pr. 13
\end{center}

\begin{center}
{\bf Abstract}
\end{center}

\vskip 0.1cm

I review several scenarios of magnetar formation in binary systems via
spin-up of a progenitor due to interaction with its companion. Mostly, these
evolutionary channels lead to formation of isolated magnetars, and indeed,
all well-established sources of this class are single objects. However, some
binaries can survive, and several candidates to accreting magnetars have
been proposed. I discuss this issue, and conclude that new accretion models
can explain properties of the proposed candidates without large magnetic
field in correspondence with models of magnetic field evolution. 


\section{Introduction}

Magnetars appear as two types of known sources: anomalous X-ray pulsars
(AXPs) and soft gamma-ray repeators (SGRs). 
Now it is well-established that magnetars are not very rare and peculiar neutron
stars (NSs). A significant fraction of NSs can be born
as magnetars. So, these objects are an important part of the general picture of
formation and evolution of compact objects.  
Still, it is not known what determines if a
newborn NS is a magnetar or not. Several proposals have been made. For
example, it was proposed
that magnetars might have very massive progenitors (\citealt{muno2007}).
Or, that magnetars have progenitors with large magnetic field
(\citealt{fw2006}).
However, observational evidence (see below)
and theoretical considerations (see \citealt{spruit2008} for a critical
discussion of the flux conservation model of magnetar formation) do not support the
universality of such scenarios (which does not exclude that they can be
applicable in some cases).  

In this note, based on a talk presented in September 2014 at the conference
``Ioffe Workshop on GRBs and other
transient sources: Twenty Years of Konus-Wind Experiment'' in St. Petersburg
(Russia), I briefly review different proposals related to evolution of
magnetar progenitors in binary systems. In addition, I discuss magnetar
evolution in accreting X-ray binary systems, and comment on related issues. 

\section{Observational evidence for magnetar origin in binaries}

 There is evidence that magnetars can originate from close
binary systems. The first indirect indication was presented by \cite{davies2009}.
These authors analysed data on SGR 1900+14 and discussed several realistic
scenarios of formation of this source in a binary system. However, the main argument
appeared very recently (\citealt{clark2014}).

 \cite{clark2014} studied the magnetar CXOU J1647-45.  It was proposed
before that this NS might have a very massive progenitor (\citealt{muno2006}). 
\cite{clark2014} observed a massive runaway star in the young cluster
Westerlund 1. Analysis of its history demonstrated that it might be born
in a binary system, disrupted after a SN explosion. The runaway star is
the former primary (initially more massive) component of the binary, the
secondary component became a progenitor of the magnetar CXOU J1647-45.   

 In this scenario the important point is that the progenitor of the magnetar
enhanced its angular momentum during evolution in the binary. This can be an
important feature which allows to understand what is special about magnetar
formation. And it confirms the main prediction of theoretical scenarios
of magnetar formation in binaries, proposed several years ago.

\section{Theoretical scenarios of magnetar formation in binaries}

 Before the first observational evidence of the origin of magnetars in
binary systems appeared, we have
studied possible evolutionary channels for magnetar
formation in binaries (\citealt{pp2006, bp2009}). 

The main motivation for our first study was a little bit peculiar. 
Even up to now all well-established magnetars (SGRs and AXPs) are isolated
objects (see the catalogue in \citealt{ok2014}). 
Most of massive stars are born in binaries, and about 10\%
of NSs remain in binary systems after supernova (SN) explosions. 
So, it seemed
reasonable to look at evolutionary channels which, on one hand results
mostly in
single NSs, and on the other hand --- maximize the probability of magnetar
formation.

In the original magnetar model by \cite{dt1993} it is proposed that huge
magnetic fields are generated by a dynamo mechanism. So, rapid initial
rotation is necessary. However, stellar cores can efficiently loose angular
momentum when a star enters the red (super)giant stage if magnetic fields
contribute to the angular momentum redistribution (e.g.
\citealt{langer2012}, and a recent review \citealt{yoon2015}). 
Thus, it is necessary to enhance rotation. 
The only effective way to do it --- is to transfer angular
momentum in a binary system (substellar components also can be in the game). 
In our studies we used population synthesis (\citealt{ufn2007}) to identify
evolutionary tracks of binary systems which allows to enhance rotation of NS
progenitors. Then we calculated the fraction of such systems, and the
fraction of isolated NSs formed through such channels.

 Note, that a similar problem was discussed in the context of 
gamma-ray burst (GRB) progenitors (see a review and references in
\citealt{yoon2015}). 
In this phenomena rotation also can play a crucial role. Then potentially,
evolutionary channels for magnetar formation and for GRB progenitors can be
similar, just in the case of GRBs a black hole (BH) formation might be required
(magnetars can also power GRBs, see a review of mechanisms in
\citealt{woosley2011}).

In addition, several studies indicated that some magnetars can have very
massive progenitors (\citealt{figer2005, muno2006}). 
Potentially, it can also fit the binary evolution
scenario, as mass loss in a binary can result in smaller core mass, and so
it opens the possibility to form a NS from initially very massive progenitor, which
otherwise might produce a BH.  However, we did not study such evolutionary
channels.

\subsection{Optimistic scenario of magnetar formation is binaries}

 In the first study (\citealt{pp2006}) we made an optimistic assumption, 
that after rotation of a progenitor is
enhanced it is not significantly reduced. I.e. that a NS can increase its magnetic
field due to a dynamo mechanism if the progenitor gained angular momentum at
any stage of its evolution. 

There are the following main possibilities to spin-up a star in a binary:
\begin{itemize}
\item merger;
\item accretion mass transfer;
\item tidal synchronization.
\end{itemize}
We analysed all of them. 

 We obtained that the fraction of NS progenitors with enhanced rotation is
larger than the expected fraction of magnetars among all NSs. This is good
news, as clearly not all NSs born from such progenitors are necessarily
magnetars. 

 We also obtained that it is easy to reproduce the large fraction of
isolated magnetars. The main channels for magnetar formation in our study
are coalescences and spin-up of the secondary companion (mostly, binaries
are destroyed after the second explosion: the fraction of survivers is low
enough to be consistent with the present-day magnetar statistics). 
It is interesting to note, that if a magnetar remains in a  binary after the
SN, then the most probable companion in this scenario is a BH.

Many magnetars in this scenario (as well as in the pessimistic scenarion,
see below) are expected to be born after SN Ib/c (see \citealt{yoon2015} for
discussion of SN Ib/c progenitors, especially in binaries). Also it is
important to note that in this evolutionary channel NSs can be formed from
low-mass stars, which can leave a white dwarf if they evolve in isolation.
 
\subsection{Pessimistic scenario}

 To be on the safe side we also studied another scenario (\citealt{bp2009}). 
Here we assumed that a star might be spun-up after the helium core
formation in a very close binary due to tidal synchronization. 
This guarantees that the angular momentum is not lost before the SN. 
 Our analysis demonstrated that we are interested only in systems with orbital period
less than $\sim 10$~days at the moment of synchronization, because only 
they can produce  NSs rotating rapidly enough.  
Of course, the fraction of NSs which passes through such a channel is smaller than in
the optimistic scenario. However, it is still consistent with magnetar
statistics, i.e. the total number of NSs born from such progenitors can
(barely) fit the number of magnetars. 

The problem is with solitarity of magnetars. Most of NSs spun-up
in such close systems remains in binaries after a SN explosion. To
reproduce the fraction of isolated magnetars we have to assume that kick is
larger for such objects than for the whole NS population. Typical values
might be above $\sim(500-700)$~km~s$^{-1}$ depending on additional
assumptions (velocity distribution, correlation with spin axis, etc.) 

Observations of magnetar spatial velocities do not support the hypothesis
that these sources have significantly larger velocities than other NSs
(see \citealt{tck2013} and references therein). 
 So, we do not think that this pessimistic scenario is very promising.
However, this channel can contribute to the magnetar population. Mostly,
magnetars formed through such channel remain in compact binaries. 
The most probable
candidates are main sequence stars and, again, BHs.  

\section{Magnetars in binaries}

 At the moment no classical magnetars (i.e., SGRs or AXPs) are known to be
members of a binary. Obviously, nothing can prevent detection of an SGR flare in a
binary, but no such cases are known. Also it is important to note that known
SGRs are mostly very young objects, and it is very unlikely that the
secondary companion in the binary could previously significantly influence
properties of the magnetosphere or/and crust of the magnetar, 
so that its activity terminated due to mass transfer. 
On the other hand, there are several claims that few
NSs in accreting X-ray binaries can have large magnetic field.

 Determination of magnetic fields of accreting NSs are mostly based on
indirect model dependent methods related to spin-up or/and spin-down (unless
cyclotron lines are detected), see \cite{rm2014}. 
Below we discuss several subjects related to
the problem of magnetars in binaries.

\subsection{Field decay and HMXBs}

 X-ray binaries are mainly divided into two wide classes: low-mass X-ray
binaries (LMXBs) and high-mass X-ray binaries (HMXBs). On average LMXBs are
old systems. So, NSs in them are significantly evolved, and their magnetic
fields might be decayed. In addition, strong accretion (typical for LMXBs)
can result in strong decrease of magnetic fields, and we observe it in
millisecond pulsars. Then we can focus on HMXBs.

 During last 35-40 years many times different authors tried to estimate
magnetic fields of accreting NSs using approaches similar (or identical) to
those described by \cite{dfp1979} (see, for example, \citealt{lp2001} and
references therein). In many cases estimates of magnetic fields produces
very large (magnetar scale) values.
However, such estimates depend on the model of accretion (see for example
\citealt{klusho2014}, where the authors under certain assumptions obtained
large field values for many NSs in Be/X-ray binaries). 
Recently, a new
one was proposed for quasispherical accretion with significantly
subEddington rates (\citealt{spkh2012}).    

 Comparison of the new Shakura, Postnov et al. model with old ones was made
by \cite{cp2012}. Several tens of Be/X-ray binaries in the SMC were used to
obtain distributions of NS magnetic fields calculated with different
approaches.   Results have been compared with predictions of the field decay
model developed by \cite{pons2009}. In this model magnetar fields decay down
to $\sim$few$\times 10^{13}$~G on the time scale $\sim$few$\times 10^5$~yrs.

 Analysis showed that only in the case of Shakura, Postnov et al. model
there are no high magnetic fields, as expected from the model by Pons et al.
As significant field decay in magnetars on the time scale of several million
years look inevitable, we conclude that at the moment the model by Shakura,
Postnov et al. in the best description of low-rate accretion in wind-fed
X-ray binaries.

\subsection{SXP 1062}

 Interestingly, it is possible to identify a former magnetar in a  binary
system. The source SXP 1062 in the SMC was proposed as an example
(\citealt{pt2012}). 

SXP 1062 is a unique system because the age of the NS is known thanks to
discovery of a SNR related to the source (\citealt{hb2012, h2012}). 
As it is young (age $\sim(2-4)\times10^4$~yrs) --- there is a possibility to
reconstruct its history. Our study demostrated that either initial spin
period was uncomfortably long ($\sim 1$~sec)\footnote{This possibility was
initially proposed by Haberl et al. (2012).}, or the NS initially had much larger
magnetic field, $B\sim10^{14}$~G, which then decayed. 
This allows the NS to start to accrete
matter form the companion's stellar wind after reasonably short interval of
time. The present day field was estimated accoding to the Shakura, Postnov
et al. model (see \citealt{spkh2012} and \citealt{spkh2013}), 
and it is well below the magnetar value.\footnote{Note, that
estimates based on older theories provide magnetar-scale field.}

\subsection{Other magnetar candidates in binaries}

 Several other sources were also proposed to contain accreting magnetars.
The list includes GX1+4, 4U 2206+54 (\citealt{reig2012}), 
Swift J045106.8-694803 (\citealt{klus2013}), and some others.
Several of the have been discussed by \cite{pskh2014}. 
In the new model of settling accretion angular momentum is more effectively
transported outwards from a NS due to convection in the envelope. 
So, for example, the
equilibrium period (for the same value of the magnetic field) in the model
by Shakura, Postnov et al. is longer than in earlier models.

\cite{pskh2014}
conclude that in the framework of the new settling accretion model
properties of the observed sources can be explained with standard magnetic
fields. I.e., a strong assumption of huge magnetic fields survival is not
necessary. This is in correspondence with our studies (\citealt{cp2012,
pt2012}).

\section{Hidden magnetars}

 The term was coined long ago by \cite{gpz1999} to label the case when the
magnetar field is overwhelmed by plasma after a SN explosion. 
We write ``long'' because just very
recently the idea of submerged field became very popular in NS
 studies (\citealt{h2011,vp2012}). A NS for several thousand or tens of
thousand years can look as a weakly magnetized source because its field has
been covered (screened) by significant amount of matter (comparable with the
crustal mass -- $\sim 10^{-3}\, M_\odot$) accreted during a fall-back
episode. 

\cite{sl2012} demonstrated that the NS in SNR Kes 79 can be a ``hidden''
magnetar, as it has a very large (64\%) pulse fraction. Recently,
we analysed properties of the NS in SNR RCW103 (\citealt{pkk2015}),
and concluded that it also can be a ``hidden'' magnetar as the flux
of this source changes in the way compatible with a magnetar activity due to
additional heat release in the crust.  

Potentially, a compact object in the SN1987A can be a ``hidden'' magnetar. 
Now it is confirmed that the progenitor was a member of a binary, which
merged not long before the SN explosion (\citealt{mp2007}). 
This perfectly fits our scenario
for magnetar formation in binaries. Thus, as no traces of the presence of
a magnetar is noticed, we can conclude that the source is a ``hidden'' one. 

\section{Discussion}

\subsection{Alternatives to spinning up the core}

In Sec. 2 we presented two scenarios of magnetar formation in binary
systems where rotation of a progenitor has been enhanced due to interaction
with a companion.  However, binary evolution is not just the way to spin-up
a star, it can also reduce spinning down of a NS progenitor.
Evolution in a binary system can reduce the duration of the
red (super)giant stage (see \citealt{mm2014} about spin-down of a stellar
core on this stage), 
so that angular momentum losses by a stellar core
are smaller (see, for example, discussion in \citealt{clark2014}).  
This adds new possibilities to form a magnetar in a binary system.

\subsection{Variety of evolutionary channels}

 It looks quite natural that evolution in a  binary can lead to relatively
rapid rotation of the core of a progenitor.  It is known that some evolved
stars, including Ib supergiants, 
have rapid rotation (see \citealt{silva2015} and references therein),
which can be due to merging, or due to ``consumption'' of a substellar companion.
Still, it is possible that a core can be spin-up just before a SN explosion.   
Such model have been recently studied by \cite{fuller2015}.

These authors demonstrated that in some cases internal gravity waves can
spin-up a stellar core already at the red supergiant stage. Thus, this can
result in formation of NSs with millisecond spin periods, and potentially
--- in a magnetar formation via a dynamo mechanism. Also this mechanism can
explain magnetars in wide binaries, like SXP 1062. 

\subsection{GUNS and magnetars in binaries}

Most of massive stars are born in binaries.
So, most of NSs are born from members (may be former) of binary systems.  
In many cases evolution in a binary could influence properties of the
progenitor, and so --- play a role in determining initial parameters of a
NS. Then, in its final form the Grand Unification of NSs,\footnote{The term
was proposed by \cite{kaspi2010}.} --- GUNS (see, for example,
\citealt{guns2014}), --- might include ingredients related to binary
evolution. 

Up to now known NSs in binary systems do not show systematic deviations in
their initial properties (increased masses and decreased magnetic fields in
millisecond pulsars are
due to long term evolution of NS in accreting binary systems). But this can
be a selection effect. For example, by definition, NSs in close binaries are
not products of mergers. Or initial parameters are ``forgotten'' as we
observe evolved and relatively old systems. 

Studies of NS properties in HMXBs can help to probe evolution of compact
objects on the time scale from few to tens million years, and first steps are
already done. More young binary systems (especially with known ages) also
can help a lot in determining initial parameters and evolutionary laws of
NSs.   

In particular, studies of possible (may be former) magnetars in binary
systems can shed light on the origin of these objects. Diversity of
observational data favours different channels of magnetar birth. 
For example, SXP
1062, if indeed it is an evolved magnetar originated from the primary
component of the binary, tells us that spin-up of a progenitor is not a
necessary condition to form a magnetar. Oppositely,  CXOU J1647-45 favours
the scenario studied by \cite{pp2006}. 
Clearly, we need more data to understand the role of binary systems in
producing different types of NSs.

\section{Conclusions}

 There are $\sim30$ magnetars, including candidates (\citealt{ok2014}). All
of them are isolated sources. On other hand, in the standard paradigm 
of the dynamo mechanism field enhancement, it is favourable for a magnetar
to be born in a binary system from a companion which has been spun-up.
Then, the main channels of magnetar formation are either related to mergers,
or to the system disruption. It is not easy to prove this scenario, but
observations start to support this model.
 
In some cases the system can survive, and the magnetar can be
observed in an X-ray binary. However, as large magnetic fields rapidly decay
down to standard values, it is very improbable to find an active  magnetar
in an X-ray binary. Most probably, we find an evolved magnetar, and it is
non-trivial to reconstruct its evolution to prove that before it has been a
stongly magnetized compact object. 
 
Understanding of the role of binary systems in formation of magnetars (and
may be in influencing parameters of some other types of compact objects) 
can be crucially important in understanding
initial properties of NSs.

\section*{Acknowledgements}
 I thank the Organizers of the conference ``Ioffe Workshop on GRBs and other
transient sources: Twenty Years of Konus-Wind Experiment''.
 I am in debt to my co-authors of the papers results from which are
presented in this  article: 
Alexei Boromazov, Anna Chashkina, Mikhail Prokhorov, and Roberto Turolla.
Also I want to thank Andrei Igoshev, Aleksander Kaminker, and Aleksander
Kaurov for fruitful collaboration.
This research was supported by the Russian Science Foundation, project
14-12-00146. The author is the ``Dynasty'' foundation fellow.

\bibliographystyle{apj}
\bibliography{bib_binmag}

\begin{thebibliography}{41}
\expandafter\ifx\csname natexlab\endcsname\relax\def\natexlab#1{#1}\fi

\bibitem[{{Bogomazov} \& {Popov}(2009)}]{bp2009}
{Bogomazov}, A.~I. \& {Popov}, S.~B. 2009, Astronomy Reports, 53, 325

\bibitem[{{Chashkina} \& {Popov}(2012)}]{cp2012}
{Chashkina}, A. \& {Popov}, S.~B. 2012, \na, 17, 594

\bibitem[{{Clark} {et~al.}(2014){Clark}, {Ritchie}, {Najarro}, {Langer}, \&
  {Negueruela}}]{clark2014}
{Clark}, J.~S., {Ritchie}, B.~W., {Najarro}, F., {Langer}, N., \& {Negueruela},
  I. 2014, \aap, 565, A90

\bibitem[{{Davies} {et~al.}(2009){Davies}, {Figer}, {Kudritzki}, {Trombley},
  {Kouveliotou}, \& {Wachter}}]{davies2009}
{Davies}, B., {Figer}, D.~F., {Kudritzki}, R.-P., {Trombley}, C.,
  {Kouveliotou}, C., \& {Wachter}, S. 2009, \apj, 707, 844

\bibitem[{{Davies} {et~al.}(1979){Davies}, {Fabian}, \& {Pringle}}]{dfp1979}
{Davies}, R.~E., {Fabian}, A.~C., \& {Pringle}, J.~E. 1979, \mnras, 186, 779

\bibitem[{{Ferrario} \& {Wickramasinghe}(2006)}]{fw2006}
{Ferrario}, L. \& {Wickramasinghe}, D. 2006, \mnras, 367, 1323

\bibitem[{{Figer} {et~al.}(2005){Figer}, {Najarro}, {Geballe}, {Blum}, \&
  {Kudritzki}}]{figer2005}
{Figer}, D.~F., {Najarro}, F., {Geballe}, T.~R., {Blum}, R.~D., \& {Kudritzki},
  R.~P. 2005, \apjl, 622, L49

\bibitem[{{Fuller} {et~al.}(2015){Fuller}, {Cantiello}, {Lecoanet}, \&
  {Quataert}}]{fuller2015}
{Fuller}, J., {Cantiello}, M., {Lecoanet}, D., \& {Quataert}, E. 2015, ArXiv
  e-prints 1502.07779

\bibitem[{{Geppert} {et~al.}(1999){Geppert}, {Page}, \& {Zannias}}]{gpz1999}
{Geppert}, U., {Page}, D., \& {Zannias}, T. 1999, \aap, 345, 847

\bibitem[{{Haberl} {et~al.}(2012){Haberl}, {Sturm}, {Filipovi{\'c}}, {Pietsch},
  \& {Crawford}}]{h2012}
{Haberl}, F., {Sturm}, R., {Filipovi{\'c}}, M.~D., {Pietsch}, W., \&
  {Crawford}, E.~J. 2012, \aap, 537, L1

\bibitem[{{H{\'e}nault-Brunet} {et~al.}(2012){H{\'e}nault-Brunet}, {Oskinova},
  {Guerrero}, {Sun}, {Chu}, {Evans}, {Gallagher}, {Gruendl}, \&
  {Reyes-Iturbide}}]{hb2012}
{H{\'e}nault-Brunet}, V., {Oskinova}, L.~M., {Guerrero}, M.~A., {Sun}, W.,
  {Chu}, Y.-H., {Evans}, C.~J., {Gallagher}, III, J.~S., {Gruendl}, R.~A., \&
  {Reyes-Iturbide}, J. 2012, \mnras, 420, L13

\bibitem[{{Ho}(2011)}]{h2011}
{Ho}, W.~C.~G. 2011, \mnras, 414, 2567

\bibitem[{{Igoshev} {et~al.}(2014){Igoshev}, {Popov}, \& {Turolla}}]{guns2014}
{Igoshev}, A.~P., {Popov}, S.~B., \& {Turolla}, R. 2014, Astronomische
  Nachrichten, 335, 262

\bibitem[{{Kaspi}(2010)}]{kaspi2010}
{Kaspi}, V.~M. 2010, Proceedings of the National Academy of Science, 107, 7147

\bibitem[{{Klus} {et~al.}(2013){Klus}, {Bartlett}, {Bird}, {Coe}, {Corbet}, \&
  {Udalski}}]{klus2013}
{Klus}, H., {Bartlett}, E.~S., {Bird}, A.~J., {Coe}, M., {Corbet}, R.~H.~D., \&
  {Udalski}, A. 2013, \mnras, 428, 3607

\bibitem[{{Klus} {et~al.}(2014){Klus}, {Ho}, {Coe}, {Corbet}, \&
  {Townsend}}]{klusho2014}
{Klus}, H., {Ho}, W.~C.~G., {Coe}, M.~J., {Corbet}, R.~H.~D., \& {Townsend},
  L.~J. 2014, \mnras, 437, 3863

\bibitem[{{Langer}(2012)}]{langer2012}
{Langer}, N. 2012, \araa, 50, 107

\bibitem[{{Lipunov} \& {Popov}(2001)}]{lp2001}
{Lipunov}, V.~M. \& {Popov}, S.~B. 2001, Astronomical and Astrophysical
  Transactions, 19, 859

\bibitem[{{Maeder} \& {Meynet}(2014)}]{mm2014}
{Maeder}, A. \& {Meynet}, G. 2014, \apj, 793, 123

\bibitem[{{Morris} \& {Podsiadlowski}(2007)}]{mp2007}
{Morris}, T. \& {Podsiadlowski}, P. 2007, Science, 315, 1103

\bibitem[{{Muno}(2007)}]{muno2007}
{Muno}, M.~P. 2007, in American Institute of Physics Conference Series, Vol.
  924, The Multicolored Landscape of Compact Objects and Their Explosive
  Origins, ed. T.~{di Salvo}, G.~L. {Israel}, L.~{Piersant}, L.~{Burderi},
  G.~{Matt}, A.~{Tornambe}, \& M.~T. {Menna}, 166--173

\bibitem[{{Muno} {et~al.}(2006){Muno}, {Clark}, {Crowther}, {Dougherty}, {de
  Grijs}, {Law}, {McMillan}, {Morris}, {Negueruela}, {Pooley}, {Portegies
  Zwart}, \& {Yusef-Zadeh}}]{muno2006}
{Muno}, M.~P., {Clark}, J.~S., {Crowther}, P.~A., {Dougherty}, S.~M., {de
  Grijs}, R., {Law}, C., {McMillan}, S.~L.~W., {Morris}, M.~R., {Negueruela},
  I., {Pooley}, D., {Portegies Zwart}, S., \& {Yusef-Zadeh}, F. 2006, \apjl,
  636, L41

\bibitem[{{Olausen} \& {Kaspi}(2014)}]{ok2014}
{Olausen}, S.~A. \& {Kaspi}, V.~M. 2014, \apjs, 212, 6

\bibitem[{{Pons} {et~al.}(2009){Pons}, {Miralles}, \& {Geppert}}]{pons2009}
{Pons}, J.~A., {Miralles}, J.~A., \& {Geppert}, U. 2009, \aap, 496, 207

\bibitem[{{Popov} {et~al.}(2015){Popov}, {Kaurov}, \& {Kaminker}}]{pkk2015}
{Popov}, S.~B., {Kaurov}, A.~A., \& {Kaminker}, A.~D. 2015, \pasa, 32, 18

\bibitem[{{Popov} \& {Prokhorov}(2006)}]{pp2006}
{Popov}, S.~B. \& {Prokhorov}, M.~E. 2006, \mnras, 367, 732

\bibitem[{{Popov} \& {Prokhorov}(2007)}]{ufn2007}
---. 2007, Physics Uspekhi, 50, 1123

\bibitem[{{Popov} \& {Turolla}(2012)}]{pt2012}
{Popov}, S.~B. \& {Turolla}, R. 2012, \apss, 341, 457

\bibitem[{{Postnov} {et~al.}(2014){Postnov}, {Shakura}, {Kochetkova}, \&
  {Hjalmarsdotter}}]{pskh2014}
{Postnov}, K.~A., {Shakura}, N.~I., {Kochetkova}, A.~Y., \& {Hjalmarsdotter},
  L. 2014, in European Physical Journal Web of Conferences, Vol.~64, European
  Physical Journal Web of Conferences, 2002

\bibitem[{{Reig} {et~al.}(2012){Reig}, {Torrej{\'o}n}, \& {Blay}}]{reig2012}
{Reig}, P., {Torrej{\'o}n}, J.~M., \& {Blay}, P. 2012, \mnras, 425, 595

\bibitem[{{Revnivtsev} \& {Mereghetti}(2014)}]{rm2014}
{Revnivtsev}, M. \& {Mereghetti}, S. 2014, \ssr

\bibitem[{{Rodrigues da Silva} {et~al.}(2015){Rodrigues da Silva}, {Canto
  Martins}, \& {De Medeiros}}]{silva2015}
{Rodrigues da Silva}, R., {Canto Martins}, B.~L., \& {De Medeiros}, J.~R. 2015,
  \apj, 801, 54

\bibitem[{{Shabaltas} \& {Lai}(2012)}]{sl2012}
{Shabaltas}, N. \& {Lai}, D. 2012, \apj, 748, 148

\bibitem[{{Shakura} {et~al.}(2012){Shakura}, {Postnov}, {Kochetkova}, \&
  {Hjalmarsdotter}}]{spkh2012}
{Shakura}, N., {Postnov}, K., {Kochetkova}, A., \& {Hjalmarsdotter}, L. 2012,
  \mnras, 420, 216

\bibitem[{{Shakura} {et~al.}(2013){Shakura}, {Postnov}, {Kochetkova}, \&
  {Hjalmarsdotter}}]{spkh2013}
{Shakura}, N.~I., {Postnov}, K.~A., {Kochetkova}, A.~Y., \& {Hjalmarsdotter},
  L. 2013, Physics Uspekhi, 56, 321

\bibitem[{{Spruit}(2008)}]{spruit2008}
{Spruit}, H.~C. 2008, in American Institute of Physics Conference Series, Vol.
  983, 40 Years of Pulsars: Millisecond Pulsars, Magnetars and More, ed.
  C.~{Bassa}, Z.~{Wang}, A.~{Cumming}, \& V.~M. {Kaspi}, 391--398

\bibitem[{{Tendulkar} {et~al.}(2013){Tendulkar}, {Cameron}, \&
  {Kulkarni}}]{tck2013}
{Tendulkar}, S.~P., {Cameron}, P.~B., \& {Kulkarni}, S.~R. 2013, \apj, 772, 31

\bibitem[{{Thompson} \& {Duncan}(1993)}]{dt1993}
{Thompson}, C. \& {Duncan}, R.~C. 1993, \apj, 408, 194

\bibitem[{{Vigan{\`o}} \& {Pons}(2012)}]{vp2012}
{Vigan{\`o}}, D. \& {Pons}, J.~A. 2012, \mnras, 425, 2487

\bibitem[{{Woosley}(2011)}]{woosley2011}
{Woosley}, S.~E. 2011, ArXiv e-prints 1105.4193

\bibitem[{{Yoon}(2015)}]{yoon2015}
{Yoon}, S.-C. 2015, \pasa, 32, 15

\end{thebibliography}

\end{document}